\documentclass[twocolumn]{jpsj3} 
\usepackage{txfonts}
\usepackage{times}
\usepackage{color}
\usepackage{graphicx}
    \usepackage{bm}
    \newcommand{\en}[1]{\epsilon _{#1}}

    \newcommand{\Gr}[2]{\mathcal{#1}_{#2}}
    \newcommand{\pri}{^{\prime}}

    \newcommand{\dcs}[1]{c^{\dagger}_{#1 \sigma}}    			
    \newcommand{\cs}[1]{c_{#1 \sigma}}				    		
    
    \newcommand{\dc}[1]{c^{\dagger}_{#1 }}    			
    \newcommand{\cc}[1]{c_{#1 }}				    		

\title{
Repulsive interaction helps
superconductivity in fullerides
}
\author{Satoshi \textsc{Yamazaki}\thanks{E-mail: yamazaki@cmpt.phys.tohoku.ac.jp}, and Yoshio \textsc{Kuramoto}}

\inst{Department of Physics, Tohoku University, Sendai 980-8578}

\recdate{\today}
\abst{
A repulsive interaction model of superconductivity (SC) is studied 
for tight-binding models with three-fold degenerate molecular orbitals.
Taking a weak-coupling approach,  we derive dimensionless coupling 
constants for various symmetries of SC pairs.  
In addition to anisotropic SC pairs,  the s-wave pairing ($A_{g}$) can also be formed.
With the purely repulsive interaction, however, the $A_{g}$ pair is not the most stable in both bcc and fcc lattices.
The most stable SC pair for the bcc lattice has the 
$T_{g}$ symmetry, which is favored by a strongly nesting 
Fermi surface.  
In the fcc lattice, various SC symmetries have comparable coupling strengths.
With the electron-phonon interaction combined, it is likely that the $A_{g}$ pair becomes the most stable.

}

\kword{
fullerides, superconductivity, multiband, Coulomb interaction
}

\begin{document}
\maketitle
\section{Introduction}

Alkali metal doped fullerides (A$_3$C$_{60}$) show not only superconductivity with high transition temperature $T_c$ up to about $40$K, but also antiferromagnetism with A=Cs.
It is known that Cs$_{3}$C$_{60}$ can take two different cubic structures; A15 (bcc like) and fcc lattices\cite{takabayashi2009}.
The conduction bands of A$_3$C$_{60}$ are composed of 
three-fold degenerate 
$t_{1u}$ molecular orbitals of a C$_{60}$ \cite{gelfand1992}. 
Photoemission spectroscopy estimates the Coulomb repulsive interaction $U$ in the range of 1-1.5 eV\cite{lof1992}.
This magnitude of $U$ is substantially larger than the band width $W \ (\sim 0.5$ eV) \cite{gelfand1992}.
These estimates suggest that A$_3$C$_{60}$ is a strongly correlated system.  
In fact Cs$_{3}$C$_{60}$ is a Mott insulator at ambient pressure
\cite{ihara2011}, while 
other A$_3$C$_{60}$ with A = K, Rb
are superconductors at low temperature
\cite{fleming1991,rosseinsky1991,tanigaki1991}.
The insulating state of Cs$_{3}$C$_{60}$ changes to superconducting (SC) state under applied pressure, which is in line with 
other organic superconductor\cite{Lefebyre2000}.

It has been discussed that 
the SC state of A$_3$C$_{60}$ is driven mainly by the electron-phonon interaction
\cite{gunnarson1999}. 
The dynamic Jahn-Teller effect is invoked for reducing the strong Coulomb repulsion, and thereby stabilizing the ordinary s-wave state \cite{han2003}.  
In view of nearby presence of the antiferromagnetic state, however, 
the effective Coulomb repulsion should remain significant even in superconducting state.  
Then the basic question remains why the $T_c$ in fullerides is so high.

The interplay of the Coulomb and electron-phonon interactions is a highly delicate matter in the strong-coupling region with high $T_c$.
As a possible step toward the complete understanding, 
we take in this paper a complimentary approach, and study how the Coulomb repulsion {\it helps} superconductivity under the characteristic band structure with degenerate orbitals.
We show that even with the repulsive interaction model, the s-wave pair can be formed.
The situation is similar to that discussed for iron pnictide compounds\cite{kuroki2010}.
We further discuss
(i) which symmetry is most stabilized in the presence of degenerate orbitals, and 
(ii) how the difference between the Fermi surfaces of bcc and fcc polymorphs 
influences the pairing.

Among many energy bands in A$_3$C$_{60}$, we keep only 
the $t_{1u}$ molecular orbitals located near the Fermi level.
In order to deal with the Coulomb repulsion, we follow and extend the weak-coupling theory for single orbital \cite{kohn1963,rague2010,hlubina1999,kondo2002}.
Such extension has already been performed in the literature, especially for another multiband system Na$_x$CoO$_2\cdot y$H$_2$O \cite{yanase2005}.
Our treatment is similar to that in ref.\citen{yanase2005}, but our motivation and conclusion is very different.

In the following section, we present our
model and discuss the perturbative approach in the presence of multi-orbitals.
The explicit results for the coupling constants are presented in \S 3 for both bcc and fcc lattices.  Section 4 summarizes the result, and discusses possible relevance of our results to actual fullerides.  The technical details of calculation are given in Appendices A and B.

\section{Gap equation with multi-orbitals}
We focus on the three molecular orbitals of each C$_{60}$ with 
$t_{1u}$ symmetry, and denote them as $m = x, y, z$. 
Although the actual C$_{60}$ molecules in A$_3$C$_{60}$ have two different orientations\cite{gelfand1992},
we simply assume uniform orientation of C$_{60}$ molecules.
Then the space groups in our models become $Bm3$ (bcc) and $Fm3$ (fcc).
By going to the momentum space, we introduce the annihilation (creation) operator
$\cs{\bm{k}m}$ \ ($\dcs{\bm{k}m}$) 
for conduction electrons with momentum $\bm{k}$ and spin $\sigma$.
The band structure is described by 
the effective tight-binding Hamiltonian as
\begin{align}
\mathcal{H}_0&=\sum_{\bm{k}\sigma}\sum_{mn}\en{mn}(\bm{k})\dcs{\bm{k}m}\cs{\bm{k}n}, \label{H0} \\
&=\sum_{\bm{k}\sigma}\sum_{l}\en{l}(\bm{k})\dcs{\bm{k}l}\cs{\bm{k}l}, \\
&\cs{\bm{k}l}=\sum_{m} A_{l}^{(m)}(\bm{k}) \cs{\bm{k}m},
\end{align}
where $l$ is the band index with the energy $\en{l}(\bm{k})$.  The latter
is obtained from diagonalization of the $3\times 3$ matrix $\en{mn}(\bm{k})$ 
with use of $A_{l} ^{(m)}(\bm{k})$ as transformation from orbital ($m$) basis.
Details of the hopping integrals involved in $\en{mn}(\bm{k})$ 
are explained in Appendix A.

\begin{figure}
\centering	
  \includegraphics[width=65mm]{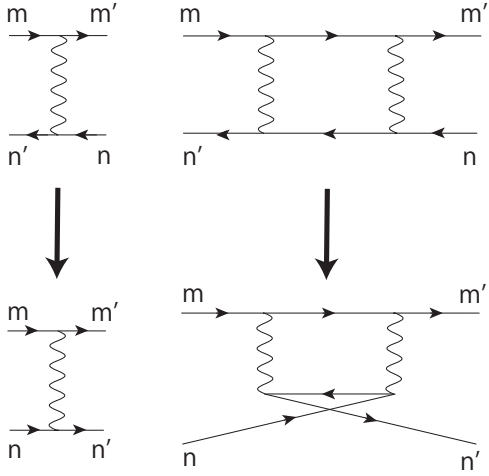}
  \caption{
Illustration of the KL interaction with $m, n, m\pri, n\pri$ being orbital indices.
The diagrams in the upper part shows 
mean $ \hat{U}$ and $\hat{U}\hat{\chi} (\bm{k}+\bm{k}\pri)\hat{U}$
in the particle-hole channel.
By exchanging the orbital labels in the lower line,
we obtain KL interaction $V_{m n ,m\pri n\pri}(k,k\pri)$ in the particle-particle channel as shown
in the lower part.
 }\label{int}
\end{figure}

We consider the on-site Coulomb interactions such as 
intra-orbital Coulomb $U$,
the inter-orbital Coulomb $U\pri$,
the Hund's coupling $J$, and
 the pair-hopping $J\pri$.
In deriving the pairing interaction originating from the Coulomb repulsion, 
we take the second-order perturbation theory according to ref.\citen{kohn1963}, and refer to it 
as the Kohn-Luttinger (KL) interaction \cite{kohn1963}. 
The KL interaction has been 
applied to several lattice models \cite{rague2010,hlubina1999,kondo2002}. 
In the orbital degenerate system, the interaction depends on orbital indices
$m, n, m\pri n\pri$, and is given by
\begin{align}
V_{m n ,m\pri n\pri}(\bm k,\bm k\pri)
=\left[ \hat{U}+\hat{U}\hat{\chi} (\bm{k}+\bm{k}\pri)\hat{U} \right]_{m n\pri , m\pri n}, \label{KL}
\end{align}
where $\hat{\chi}  (\bm{k}+\bm{k}\pri)$ is the momentum-dependent static susceptibility matrix.
The Coulomb interaction $\hat{U}$ has matrix elements:
$U_{m m, m m}=U
,\  U_{m n ,m n}=U\pri 
,\  U_{m m ,n n}=J 
,\  U_{m n ,n m}=J\pri, \ (m\neq n) $. 
The other components are set to zero. 
Figure \ref{int} shows how to label the KL interaction with orbital indices.
The interaction depends on the sum $\bm{k}+\bm{k}\pri$ of the internal momenta of incident and scattered pairs.

The matrix elements of $\hat{\chi}$ are given by 
\begin{align}
\chi _{mn,m\pri n\pri}(\bm{q})=& -\frac{1}{N}\sum_{\mib p,l,l\pri}
\bar{A}_l^{(m)}(\bm{p}+\bm{q})A_l^{(m\pri)} (\bm{p}+\bm{q})A_{l\pri} ^{(n)}(\bm{p})\bar{A}_{l\pri} ^{(n\pri)}(\bm{p}) \notag \\
&\qquad \times \  \frac{f(\en{l}(\bm{p+q}))-f(\en{l\pri}(\bm{p}))}{\en{l}(\bm{p+q})-\en{l\pri}(\bm{p})},
\label{chi}
\end{align}
where $f(\en{}) =1/(e^{\beta(\en{} -\mu)}+1)$ is the Fermi distribution function with  
$\mu$ the chemical potential and $\beta$ the inverse of temperature $T$. 
The bar in $\bar{A}_l^{(m)}$ denotes the complex conjugate, and
$N$ is number of 
lattice points in the system. 
Note that $\hat{\chi}$ sensitively reflects the shape of Fermi surface (FS), especially the nesting structure.
Namely the KL interaction becomes strong if $\bm{k}+\bm{k}\pri$ corresponds to a nesting vector, 


We derive the dimensionless SC coupling constant $\lambda$ in the following manner.
Let us start with the linearized gap equation given by
\begin{align}
\Delta_l(\bm{k})=-\frac{1}{N}\sum_{\bm{k}\pri l\pri}V_{ll\pri}(\bm{k},\bm{k}\pri)\frac{\tanh{(\en{l\pri}(\bm{k}\pri)/2T_c})}{2\en{l\pri}(\bm{k}\pri)}\Delta_{l\pri}(\bm{k}\pri), \label{gapeq}
\end{align}
where 
$\Delta_{l}(\bm{k})$ is the SC gap function with the band index $l$. 
Note that the interband pairing is not realized in the weak coupling limit.
Accordingly the pairing interaction is also labeled by the band indices as
\begin{align}
V_{ll\pri}(\bm{k},\bm{k}\pri)=&\notag\\  \sum_{m n , m \pri n \pri} \bar{A}_{l}^{(m)}(\bm{k}) &\bar{A}_{l}^{(n)}(-\bm{k})
 V_{m n , m \pri n \pri}(\bm{k},\bm{k}\pri ) A_{l\pri}^{(m \pri)}(\bm{k}\pri) A_{l\pri}^{(n \pri)}(-\bm{k}\pri). \label{band KL}
\end{align}

We work with the weak coupling limit where $\Delta_{l}(\bm{k})$ is sizable only near the FS. %
Since $T_c$ is much smaller than the characteristic energy of $\hat{\chi}$,
we can neglect the $T$-dependence of the effective interaction $V_{m n , m \pri n \pri}(\bm{k},\bm{k}\pri )$, and put $T=0$ in eq.(\ref{chi}).
Then, we can rewrite eq.(\ref{gapeq}) as
\begin{align}
\Delta_{l}(\bm{k}) & \simeq -\frac{1}{N}\sum_{\bm{k}\pri l\pri}V_{ll\pri}(\bm{k},\bm{k}\pri)
 \delta(\en{\bm{k}\pri l\pri})\Delta_{l\pri}(\bm{k}\pri) \notag \\
& \qquad \times \int^{\omega_c}_{0} d\en{}\frac{\tanh{(\en{}/2T_c})}{\en{}},
\end{align}
where $\omega_c$ is the cut-off energy.
By changing the summation over $\bm{k}$ by the surface integral over the FS,
we obtain the eigenvalue equation for $\lambda$ as 
\begin{align}
\lambda \ \Delta_{l}(\bm{k})=-\frac{1}{(2\pi)^3}\sum _{l\pri}\int _{\en{\bm{k}l\pri}=
\en{F}} \frac{dS_{\bm{k}\pri l\pri}}{v_F(\bm{k}\pri,l\pri)}V_{ll\pri}(\bm{k},\bm{k}\pri)\Delta_{l\pri}(\bm{k}\pri),\label{eigen}
\end{align}
where
$dS_{\bm{k} l}$ denotes a FS element of the band $l$ with
$v_F(\bm{k},l)$ being its Fermi velocity, and 
\begin{align}
\frac{1}{\lambda}\equiv  \int^{\omega_c}_{0} d\en{}\frac{\tanh{(\en{}/2T_c})}{\en{}} .  
\label{Tc}
\end{align}
Equation (\ref{eigen}) shows that $\lambda$ is independent of $\omega_c$, while eq.(\ref{Tc}) shows
that $T_c \sim \omega_c \exp(-1/\lambda)$ does depend on $\omega_c$.
In the following, we concentrate on $\lambda$ rather than $T_c$.
We evaluate the surface integral by using the tetrahedron method\cite{rath1975}, 
and the eigenvalue equation (\ref{eigen}) is solved by the power method. 
In our calculation, more than $10^4$ elements on FS are used.

In general, a SC order parameter involves all conduction bands with a component $\Delta_{l}(\bm{k})$ in each band $l$. 
The largest eigenvalue leads to the highest $T_c$ and its eigenvector determines the symmetry of the SC state\cite{graser2009}.
Since each C$_{60}$ molecule does not have the four-fold rotation axis, the SC state belongs to the irreducible representation of point group $T_h$ \cite{sergienko2004}.
Then, some gap functions in the cubic $O_h$ group with higher symmetry are mixed under the $T_h$ symmetry \cite{koga2006}.
For the spin singlet Cooper pair,
there are three representations: $A_g, E_g, T_g$.
Imposing these symmetries on $\Delta_{l}(\bm{k})$,
we reduce the $\bm{k}$-space integration in eq (\ref{eigen}) to a smaller part of the BZ.
In this way we obtain eigenvalue equations for each symmetry.
More details of the reduction procedure is explained in Appendix B.
In order to avoid the $O(U)$ term in the KL interaction,
a nodal SC state is favored in a single band model\cite{rague2010}. 
However, 
since $\Delta_{l}(\bm{k})$ can change the phase factor at each band,
even a fully gapped SC state can avoid the $O(U)$ term in the multiband model.
Note that similar situation has been discussed in iron pnictide compounds\cite{kuroki2010}.

\section{Numerical Results}

\subsection{Choice of parameters}
We have derived the maximum $\lambda$ for each symmetry of singlet pairing with fcc and bcc crystal structures.
Since the triplet pair is less interesting for fullerides, we have only derived $\lambda$ with $T_u$ symmetry.
In the following, 
we consider only the case of three conduction electrons per C$_{60}$.
We fix the band width $W$ to 0.5 eV.
Concerning the exchange interaction, it has been argued that the Jahn-Teller interaction can change the sign of the Hund's rule coupling \cite{capone2009}.  In view of such uncertainty, we simply put 
$J = J\pri =0$, and $U=U\pri$.
Then we obtain the simple scaling $\lambda\propto U^2$, provided only the $O(U^2)$ term in eq.(\ref{KL}) is active.
The values of $\lambda$ computed for each symmetry are summarized in Table \ref{eigentable}.  Note that the theory remains accurate only for $U$ much smaller than the band width $W$.

\begin{table}
\caption{The SC coupling constant 
$\lambda$ for each irreducible presentation (IR).    The unit of $U$ is eV, which means that $\lambda$ takes the numerical value in the table with $U=1$eV, although the relevant magnitude for $U$ is much smaller.
}
\begin{center}
\begin{tabular*}{70mm
}{cc|c|c}  \hline
IR  & Degeneracy &  $\ \lambda/U^2\ $ (bcc) & $\ \lambda/U^2\ $ (fcc) \\ \hline
$A_{g}$ &1 &  0.2358 & 0.2016  \\
$E_{g}$ &2 &  0.1349 & 0.2926  \\
$T_{g}$  &3 &  0.6287 & 0.2552  \\ \hline
$T_{u}$  &3 &  0.2860 & 0.1869  \\
 \hline
\end{tabular*}
\label{eigentable}
\end{center}	
\end{table}


\subsection{bcc model}
\begin{figure}
\centering	
  \includegraphics[width=85mm]{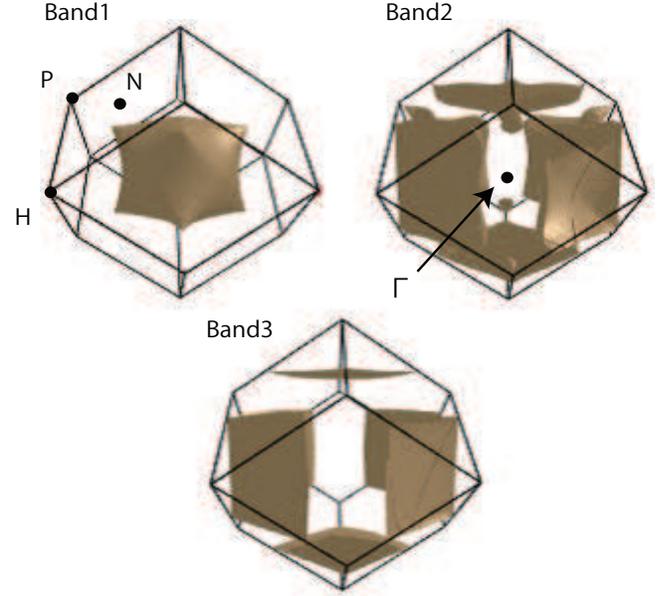}
  \caption{The FS in the the bcc model.  The center of the BZ is $\Gamma$.
See text for comment on the shape.
The FS in band 1 is shaped like a cube. FSs in band 2 and band 3 have six sheets which are almost flat. 
If we combine 6 sheets at each bands, FS in band 2 and band 3 also look like cube which is centered in H point.
In single band bcc model, FS becomes a cubic shape perfectly.
}\label{bcc-l02}
\end{figure}

We have fitted the tight-binding parameters so as to reproduce the energy bands derived by a more elaborate method \cite{erwin1991}.
With only the nearest- and next-nearest hoppings included, a satisfactory fitting can be performed, as explained in Appendix A.  
Figure \ref{bcc-bands} shows the calculated result for conduction bands in the bcc model.
\begin{figure}
\centering	
  \includegraphics[width=85mm]{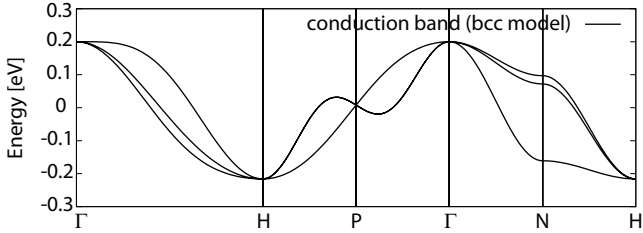}
  \caption{
The conduction bands derived in the tight-binding bcc model.  
}\label{bcc-bands}
\end{figure}%
The resultant FS is shown in Fig.\ref{bcc-l02}.
The FS in band 1 is shaped like a cube, while the FS's in band 2 and band 3 
are not closed the first BZ.  
As a result the FS consists of 
six sheets that are very flat in band 3, and less flat in band 2.
If we combine six sheets at each band beyond the first BZ, the FS in band 2 and band 3 also looks like a cube centered on H point in the BZ.
Note that a perfect cube of the FS is obtained in a half-filled single band model with only the nearest neighbor hopping in the bcc lattice.
This feature is weakened by band mixing, but remains to some extent in the present bcc model.

\begin{figure}
\centering	
  \includegraphics[width=85mm]{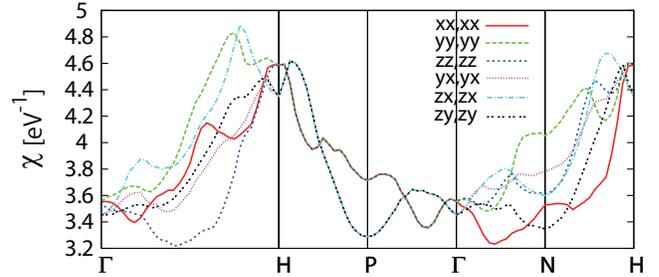}
  \caption{Momentum dependence of the main components $\chi _{m n, m n}(\mib q)
  \ (m, n =x,y, z)$.
Location of high symmetry $\mib q$ points is illustrated in Fig.\ref{bcc-l02}.
The enhancement  near $H$ point is related to antiferromagnetic fluctuation in the bcc model.  Note that summation of $\chi _{m m, n n}$ over $m$ and $n$ recovers the $T_h$ symmetry.
}\label{chiK6}
\end{figure}

The strong nesting in the FS appears in the momentum dependence of the static susceptibility $\chi (\mib q)$. 
Figure \ref{chiK6} shows the main components $\chi _{mn, mn}$.
The other components $\chi _{m m,n n}$ and $\chi _{m n, n m}$ with $m \neq n$ are 
smaller by an order of magnitude than $\chi _{mn, mn}$.
Those components with three different indices
such as $\chi_{xy,zx}$ are zero since each molecular orbital is odd under space inversion. 
In Fig.\ref{chiK6}, 
the enhanced response 
around $\mib q = \mib Q_H=(2\pi,0,0), (0, 2\pi, 0), (0,0,2\pi)$
is related to instability toward antiferromagnetic (AF) order.
This instability is strong in the bcc lattice with bipartite structure. 
In the single band model with perfect nesting at half filling,
the AF susceptibility $\chi(\mib Q_H)$ diverges at $\mib Q_H$.

\begin{figure}
\centering	
  \includegraphics[width=85mm]{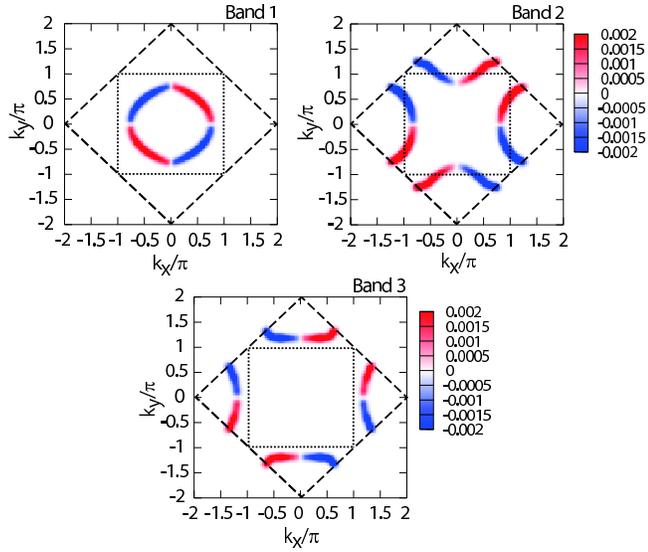}
  \caption{
Results for the gap function $\Delta_l(\bm{k})$ with $d_{xy}$ symmetry
projected onto the plane $k_z=0$
in the bcc model.
The order parameter with $T_g$ symmetry consists of three components 
$d_{xy}, d_{yz}, d_{zx}$.
The dashed lines show the boundary of bcc BZ, and the dotted line shows
the completely nesting FS of the single band model with only the nearest-neighbor hopping.
The $d_{xy}$ state has nodal points at $k_x=0$ and $k_y=0$, which extends to lines by including finite $k_z$.
}\label{bccgap}
\end{figure}

Using the susceptibility thus derived, we calculate the SC coupling constant.
The largest eigenvalue of $\lambda$ in eq.(\ref{eigen}) for each symmetry are shown in Table \ref{eigentable}.
In the bcc model, the $T_g$ symmetry of the pair is most favored.
The eigenstate has the three-fold degeneracy with components ($d_{xy}$, $d_{yz}$, $d_{zx}$). 
In order to understand the stability of the $T_g$ pairing,
we show in Fig.\ref{bccgap}  the gap function $\Delta _l(\bm{k})$
of the $d_{xy}$ state along the FS
in the $k_z=0$ plane in BZ. 
The numerical values shown by color mean the amplitude with 
normalization over the whole Fermi surface.
The red and blue parts represent positive and negative $\Delta _l(\bm{k})$, respectively, that can be taken real.  
The boundary between red part and blue part becomes a node.
For comparison the completely nested FS in the single-band case is shown by dotted lines.  
The band 2 and band 3 satisfy the nesting condition reasonably well. 

It is known in the single band bcc model\cite{rague2010}
that the SC state with $T_g$ symmetry is most stabilized. 
To understand this as well as our result for the multiple bands,
suppose in Fig.\ref{bccgap} that a SC pair with red region of $\mib k$ is scattered to $\mib k'$ in the blue region in the same band.  
With the $d_{xy}$ symmetry, there is substantial combination to satisfy
$\mib k+\mib k\pri \sim \mib Q_H$.
Hence, the effective interaction $V_{ll\pri}(\bm{k},\bm{k}\pri)$ becomes 
effective for $d_{xy}$ state.  Similar situation occurs to other members
$d_{yz}, d_{zx}$
of 
the $T_g$ symmetry.
Note that 
the gap functions of all bands have the same sign for each quadrant 
of BZ in Fig. \ref{bccgap}.
This is because each band has comparable size of FS, and compose almost degenerate bands. 

\begin{figure}
\centering	
  \includegraphics[width=85mm]{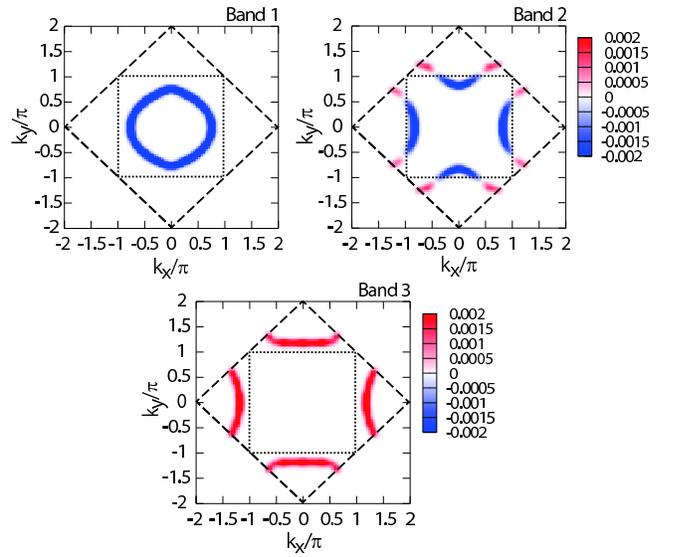}
  \caption{
Illustration of the gap function $\Delta_l(\bm{k})$ with $A_g$ symmetry
projected onto the plane $k_z=0$ in the bcc model.
Note that the gap functions of bands 1 and 3 have opposite signs, while the band 2 has nodes keeping the $A_g$ symmetry.
}\label{bccAg}
\end{figure}

We next move to the pairing with the $A_g$ symmetry which is less stable in the present model, but which seems most relevant to actual fullerides.
Figure \ref{bccAg} shows the gap function $\Delta_l(\bm{k})$ for each band. 
Different signs of the (real) gap functions work
to cancel the $O(U)$ part of the KL interaction.  
We emphasize such cancellation requires nodes along the FS in the single band model.
In our case, the band 2 has nodes as in the single-band model, while band 1 and 3 cancel the $O(U)$ without nodes, and hence having less cost in the kinetic energy.

\subsection{fcc model}
\begin{figure}
\centering	
  \includegraphics[width=80mm]{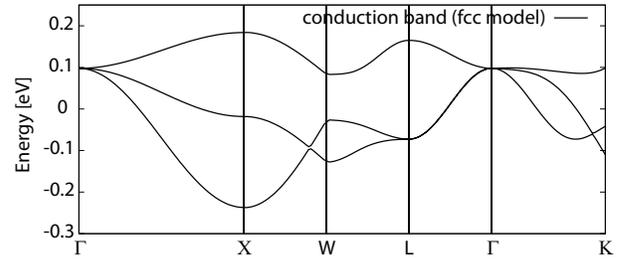}
  \caption{
The conduction bands in the tight-binding fcc model.  
}\label{fcc-bands}
\end{figure}

\begin{figure}
\centering	
  \includegraphics[width=80mm]{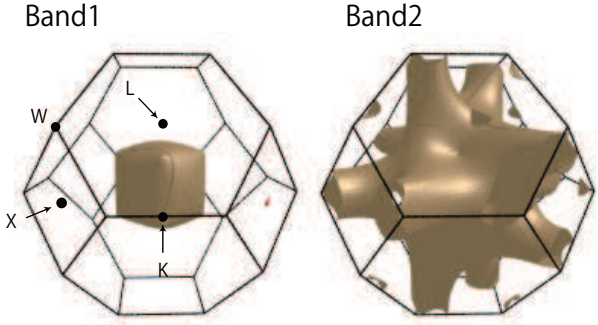}
  \caption{Two sheets of the FS in the fcc model.
High symmetry points in the first BZ are shown.  See text for comments on the shapes.   
}\label{fcc-l02}
\end{figure}

The tight-binding fit of the band structure of Cs$_3$C$_{60}$ is available in literature\cite{gelfand1992}.  We have used the same fitting parameters and reproduced the previous result. For completeness, we show our result for the fcc model in Fig.\ref{fcc-bands}.
Figure \ref{fcc-l02} shows the FS in the fcc model.
Although the conduction bands consist of three branches, there are only two FS's.
It is clear in band 2 that there is no four-fold symmetry in the FS, which is however consistent with the local $T_h$ symmetry of C$_{60}$.
Note the presence of the three-fold symmetry around the [111] axis.

\begin{figure}
\centering	
  \includegraphics[width=85mm]{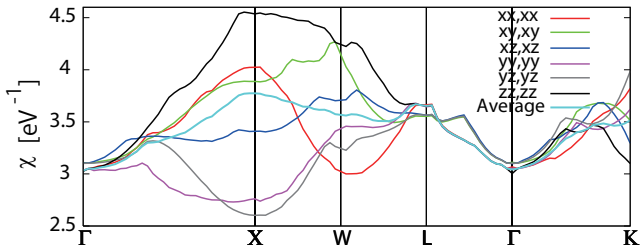}
 \caption{The momentum dependence of the main components $\chi _{m n , m n } \ (m, n=x, y,z)$ together with the average.
High symmetry points are illustrated in Fig.\ref{fcc-l02}.
From X point to W point, some $\chi _{mm ,mm}$ becomes large,
but the average of $\chi _{mm ,mm}$ has no special peak.
}\label{chiCs}
\end{figure}

Figure \ref{chiCs} shows the momentum dependence of the main components $\chi _{mn, mn}$.
The other components are negligible as in the bcc model.
The average of $\chi_{mm,mm}$ is shown by right blue line, which does not have a particular wave vector for enhanced behavior.
This corresponds to a difficulty to realize AF order in fcc lattice which has a geometrical frustration.
Using this susceptibility, we calculate the SC coupling constant.
In Table \ref{eigentable}, the SC state with $E_{g}$ symmetry is the most stable in the fcc model, but the difference with other symmetry is not large. 

\begin{figure}
\centering	
  \includegraphics[width=85mm]{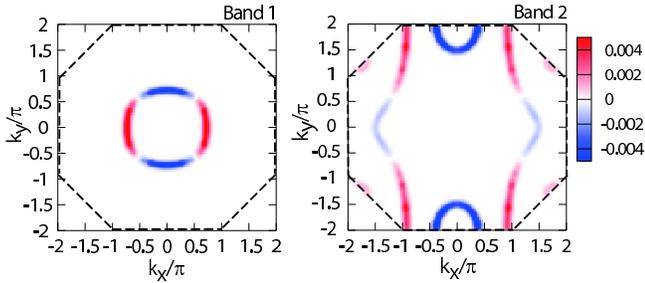}
  \caption{
Illustration of the gap function $\Delta _l(\bm{k})$ with $E_g$ symmetry
in the fcc model.
The dotted lines show the boundary of fcc BZ.  
The nodes in band 2 has no four-fold symmetry
because of the $T_h$ symmetry of each C$_{60}$ molecule.
}\label{fccgap}
\end{figure}

Figure \ref{fccgap} shows the gap function $\Delta _l(\bm{k})$ with the 
$E_g$ symmetry in the fcc model
in the $k_z=0$ plane of BZ. 
In contrast with the $O_h$ group, which has $d_{x^2-y^2}$ and $d_{3z^2-r^2}$ states as eigenfunctions,
the eigenfunctions in the $T_h$ symmetry have no four-fold symmetry.  
In our result, $\Delta _1(\bm{k})$ corresponding to band 1 is close to $d_{x^2-y^2}$ state, 
while $\Delta _2(\bm{k})$ is rather different from the $E_g$ state in the $O_h$ group.
The node structure of $\Delta _l(\bm{k})$ in $E_g$ symmetry is complicated.
In band 1, the nodal structure is rotated by $\pi/4$ from the corresponding band in the bcc case.
On the other hand, in band 2, the node appears not only along the FS but also in disconnected parts along $k_y =\pm 2\pi$ of the FS.  

\begin{figure}
\centering	
  \includegraphics[width=85mm]{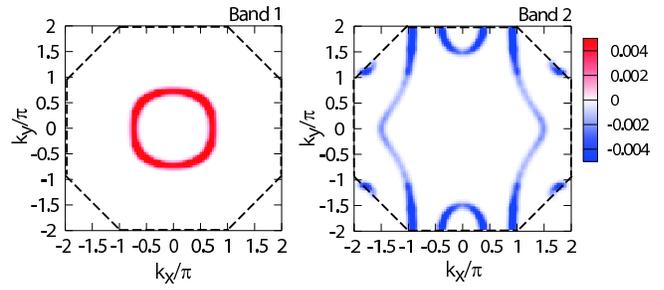}
  \caption{
Illustration of the gap function $\Delta_l(\bm{k})$ with $A_g$ symmetry
in the $k_z=0$ plane in the fcc model.
The two sheets of FS have opposite signs of $\Delta_l(\bm{k})$.
}\label{fccAg}
\end{figure}

Finally we present the results for the $A_g$ pairing in the fcc model.
The coupling constant $\lambda$ as given in Table 1 is almost the same as in the bcc model.
Figure \ref{fccAg} shows the gap function in the $k_z=0$ plane of the BZ.
There is no node for each FS in accordance with the s-wave like character.  However, the sign of $\Delta_l(\bm{k})$ is opposite between bands 1 and 2.   In this way, the first-order Coulomb repulsion is canceled.

\section{Summary and Discussion}

We have studied the repulsive interaction model for superconductivity in  A$_3$C$_{60}$. 
By using the second order perturbation theory with respect to the Coulomb repulsion, we calculate the SC coupling constant $\lambda$ for various SC symmetries and for both bcc and fcc lattices.
By the nesting property of the FS in the bcc model, the 
$T_{g}$ symmetry is the most favorable within the repulsive interaction model.
We have further shown that the stable pair with the fully symmetric $A_g$ can also be formed by purely repulsive interaction.  This stability essentially requires the presence of multiple conduction bands.

Let us now discuss possible relevance of our results to actual superconductivity in fullerenes, especially in Cs$_3$C$_{60}$ with the highest $T_c$.  The most fundamental question is why the $T_c$ is so high.  Since the system is close to the Mott transition, one might naively guess that the SC is caused by a non-phonon mechanism as in cuprate superconductors.  However, various experimental evidences point to the fully symmetric s-wave SC being realized.  In the conventional theory, the s-wave SC is unfavorable in the presence of strong Coulomb repulsion.

According to our results in this paper, we propose the following scenario for the SC in Cs$_3$C$_{60}$:  
What is responsible for the high $T_c$ is the cooperation, rather than competition, between 
the Coulomb repulsion and the Jahn-Teller phonons.
The $A_g$ pair will be most favorable for such cooperation, which is fully gapped in the fcc model.
In the bcc model, on the other hand, one of the three bands has four-fold nodes in the gap function.
Hence our repulsive interaction model predicts gapless s-wave superconductivity in the bcc system.
It is not clear whether the nodes remain in the presence of Jahn-Teller phonons. 
Theoretically, 
simultaneous account of Jahn-Teller phonons together with Coulomb repulsion 
requires a new scheme which is not available at present.
The main difficulty is that the KL interaction is highly non-local and hence is 
beyond the scope of the dynamical mean-field theory (DMFT).   
We have of course recognized that the DMFT can powerfully deal with local Coulomb correlation together with Jahn-Teller phonons \cite{han2003}.  
Various available schemes in the momentum space, on the other hand, are not reliable enough to deal with the case of strong Coulomb repulsion.
Namely, SC in fullerenes provides a challenging ground to construct a new theoretical scheme.  
We hope to contribute to further development of the theory in the near future, and test the scenario mentioned above.

\section*{Acknowledgment}
One of the authors (S.Y.) is supported by the
global COE program of the Ministry of Education, Culture, Sports, Science and
Technology, Japan (MEXT). 

\appendix
\section{Hopping integrals for $t_{1u}$ molecule orbitals}
We explain how the parameters are chosen in the 
effective tight binding models for A$_3$C$_{60}$.
Using the molecular orbitals specified by $m=x,y,z$, we consider the hopping integrals 
$t_{m n}(\alpha, \beta, \gamma)$ 
between C$_{60}$ molecules
with the directional cosines  $\alpha, \beta, \gamma$.
Then the effective tight binding Hamiltonian is written as
\begin{align}
\Gr{H}{0}&=\sum _{<i,j>}\sum _{\sigma, m n} 
t_{m n}(\alpha, \beta, \gamma) \left( \dc{im\sigma}\cc{jn\sigma} +h.c. \right)  \label{fccSK} \\
&=\sum _{\bm{k}, \sigma} 
\begin{pmatrix}
\cc{\bm{k}x\sigma} \\
\cc{\bm{k}y\sigma} \\
\cc{\bm{k}z\sigma} \\
\end{pmatrix}^{\dagger}
\begin{pmatrix}
\en{xx}(\bm{k})&\en{xy}(\bm{k})&\en{xz}(\bm{k}) \\
\en{yx}(\bm{k})&\en{yy}(\bm{k})&\en{yz}(\bm{k}) \\
\en{zx}(\bm{k})&\en{zy}(\bm{k})&\en{zz}(\bm{k}) \\
\end{pmatrix}
\begin{pmatrix}
\cc{\bm{k}x\sigma} \\
\cc{\bm{k}y\sigma} \\
\cc{\bm{k}z\sigma} \\
\end{pmatrix},\label{fcchami}
\end{align}
where $\sum_{<i,j>}$ is a summation over the nearest and the next nearest neighbor sites.
The parameters $t_{mn}(\alpha, \beta, \gamma)$ are fitted to be 
consistent with first principle band calculations\cite{gelfand1992,erwin1991}.


     \subsection{bcc model}\label{3bcc}
We keep only the the nearest-neighbor hopping
$\hat{t}_n$ and the next-nearest-neighbor one
$\hat{t}_{nn}$.
the bcc model. 
Among eight nearest neighbors and six next-nearest neighbors, we take as representative members, a nearest neighbor along $(1,1,1)$ and a next-nearest neighbor along $(1,0,0)$.
The hopping integrals are parameterized as
\begin{align*}
\hat{t}_n &=
\begin{pmatrix}
X&Y&Y \\
Y&X&Y \\
Y&Y&X \\
\end{pmatrix},
&\hat{t}_{nn} &= 
\begin{pmatrix}
Z&0&0 \\
0&W&0 \\
0&0&W\pri  \\
\end{pmatrix},
\end{align*}
where three orbitals of $t_{1u}$ have been taken as
the basis of the matrix.
The hopping integrals to other equivalent neighbors are determined by symmetry.
Fitting to a first-principle calculation \cite{erwin1991}, we choose the following values:
\begin{align}
X&=0.026,&
Y&=-0.015,& \notag \\
Z&=0.016,& 
W&=-0.008,&
W\pri &=-0.025. \label{w}
\end{align}
The inequality $W\neq W\pri$ comes from the absence of the four-fold symmetry in $T_h$, which is beyond   
the Slater-Koster parameterization for $p$ orbitals \cite{slater1954}.

Note also that actual Cs$_3$C$_{60}$ takes the A15 structure where 
C$_{60}$ molecules have different orientations in the body-center and corner positions.
Our form of $\hat{t}_n$ assumes, following ref.\citen{erwin1991}, a simplified form that C$_{60}$ molecules are all equivalent.

\subsection{fcc model}
The fitting parameters for fcc model have been introduced already by Gelfand et al\cite{gelfand1992}.
Following their results, we keep only the nearest-neighbor hopping and use the following parameters for $\hat{t}_n$ along $(110)$:
\begin{align*}
\hat{t}_n
&=
\begin{pmatrix}
X&Y&0\\
Y&X\pri &0 \\
0&0&Z 
\end{pmatrix},
\end{align*}
with 
\begin{align*}
X&=0.0083 ,&
X\pri   &=0.0336 ,&
Y&=-0.0198 ,&
Z&=-0.0191. 
\end{align*}
The inequality $X \ne X\pri$ reflects the absence of four-fold symmetry 
in each C$_{60}$ molecule.
The actual C$_{60}$ molecules in the fcc A$_3$C$_{60}$ take one of two orientations in a random way\cite{stephens1991}.   
We assume for simplicity the same C$_{60}$ orientations as in the bcc model.
\appendix 
\section{Use of Symmetry in Deriving Eigenvalues}

We describe how to utilize the symmetry of $\Delta_l(\bm{k})$ in deriving
the eigenvalue.  For an irreducible representation
$\alpha$, eq.(\ref{eigen}) takes the form:
\begin{align}
\lambda _\alpha \ \Delta_{l\alpha}(\bm{k})&=-\frac{1}{(2\pi)^3}\sum _{l\pri}\int _{\en{\bm{k}l\pri}=
\en{F}} \frac{dS_{\bm{k}\pri l\pri}}{v_F(\bm{k}\pri,l\pri)}
\Gamma^\alpha_{ll\pri}(\bm{k},\bm{k}\pri)\Delta_{l\pri \alpha}(\bm{k}\pri), \label{eigen-sym}
\end{align}
where
$\Gamma^\alpha_{ll\pri}(\bm{k},\bm{k}\pri)$ is the effective interaction. 
It is possible to reduce the range of integration to $k'_x, k'_y ,k'_z \ge 0$ by using the symmetry operation.
Table \ref{basic} shows 
the relation between each symmetry operator $\mathcal{O}$ and irreducible representations of $T_h$.
The basic symmetry operators change the wave number $\bm{k}$ and the phase factor of the gap function. 

Note that we do not necessarily use all operators for reducing the BZ integration.   For example, since
 $C_3$ and $C_3^2$ change $d_{xy}$ basis function to other basis function,
they are not utilized in the case of $T_g$.
In this paper, we choose $A_g$, $E_g^{(1)}$, and $d_{xy}$ as basis functions.
Then,  $\Gamma^\alpha_{ll\pri}(\bm{k},\bm{k}\pri)$ for each symmetry are given in terms of the KL interaction $V_{ll\pri}(\bm{k},\bm{k}\pri)$ 
by
\begin{align}
\Gamma^{A_g}_{ll\pri}(\bm{k},\bm{k}\pri)&=
[1+\sigma _x(\bm{k}\pri)]\cdot [1+\sigma _y(\bm{k}\pri)]\cdot [1+\sigma _z(\bm{k}\pri)] \notag\\
&\quad \times \ [1+C_3(\bm{k}\pri)+C^2_3(\bm{k}\pri)]V_{ll\pri}(\bm{k},\bm{k}\pri), \\
\Gamma^{E_g}_{ll\pri}(\bm{k},\bm{k}\pri)&=
[1+\sigma _x(\bm{k}\pri)]\cdot [1+\sigma _y(\bm{k}\pri)]\cdot [1+\sigma _z(\bm{k}\pri)] \notag\\
&\quad  \times \ [1+\omega C_3(\bm{k}\pri)+\omega^2 C^2_3(\bm{k}\pri)]V_{ll\pri}(\bm{k},\bm{k}\pri), \\
\Gamma^{T_g}_{ll\pri}(\bm{k},\bm{k}\pri) &=
[1-\sigma _x(\bm{k}\pri)]\cdot [1+\sigma _y(\bm{k}\pri)]\cdot [1+\sigma _z(\bm{k}\pri)] \notag \\
&\quad  \times   V_{ll\pri}(\bm{k},\bm{k}\pri),
\end{align}
with $\omega=\exp(2\pi i /3)$.

 \begin{table}
 \caption{
 Eigenvalues of the basic symmetry operation $\mathcal{O}$ for each irreducible representation of the gap function $\Delta _l$.
 Degenerate representations such as $E_g$ and $T_g$ can have diagonal basis set by proper choice given in the Table.
 Namely, $d_{xy}$,  $d_{yz}$, and $d_{zx}$ are the basis functions of $T_g$ symmetry.
 Each operator $\mathcal{O}$ also changes the wave vector $\bm{k}$ as shown in the second column, 
 and the phase factor as shown from the third columns.
For example, the operator $\sigma_x$ changes the gap function with $d_{xy}$ symmetry as:
  $\Delta _l( -k_{x}, k_{y}, k_{z})=(-1) \times \Delta _l( k_{x}, k_{y}, k_{z})$.
}
\begin{center}
\begin{tabular*}{80mm}{c|c|cccccc}  \hline
$\mathcal{O}$  & $\mathcal{O}\bm{k}$   & $A_g$&  $E_g^{(1)}$&$E_g^{(2)}$&  $d_{xy}$&  $d_{yz}$&  $d_{zx}$ \\ \hline
$E$  &$k_{x}, k_{y}, k_{z}$     &  1   &  1&  1&  1&  1    &  1  \\
$\sigma _x$ &$-k_{x}, k_{y}, k_{z}$      &  1   &  1&  1&  -1&  1    &  -1\\
$\sigma _y$ &$k_{x}, -k_{y}, k_{z}$      &  1   &  1&  1&  -1&  -1    &  1\\
$\sigma _z$ &$k_{x}, k_{y}, -k_{z}$      &  1   &  1&  1&  1&  -1    &  -1\\
$C_3$ &$k_{y}, k_{z}, k_{x}$      &  1   &  $\omega$&  $\omega ^2$    &  -&   -&  -\\
$C_3^2$ &$k_{z}, k_{x}, k_{y}$      &  1   &  $\omega ^2$    &  $\omega$ &  -&  -&  -\\
 \hline
\end{tabular*}
\label{basic}
\end{center}	
\end{table}

We obtain the maximum eigenvalue $\lambda_{\alpha}$ by using eq. (\ref{eigen-sym}) for each basis function.  
For a degenerate representation such as $T_g$, three functions such as 
$d_{xy}, d_{yz}$ and $d_{zx}$ may enter into the process of iteration in the power method.
However, we actually need only $d_{xy}$-type functions as shown below.

Let us consider a matrix element

\begin{align}
&I_{mn}\equiv \frac{1}{(2\pi)^6}\sum _{ll\pri}\int\int \frac{dS_{\bm{k} l}}{v_F(\bm{k},l)} \frac{dS_{\bm{k}\pri l\pri}}{v_F(\bm{k}\pri,l\pri)}
\notag \\ &\qquad \qquad \times \  \left( \Delta_{l\alpha}^{(m)}(\bm{k})\right)^*V_{ll\pri}(\bm{k},\bm{k}\pri)\Delta_{l\pri \alpha}^{(n)}(\bm{k}\pri), 
\end{align} 
where $m,n$ represent either $xy, yz$ or $zx$ in $T_g$.
In our multiband model, $\Delta _{l\alpha}^{(m)}(\bm{k})$ and $V_{ll\pri}(\bm{k},\bm{k}\pri)$ satisfy the relations:
\begin{align}
\sigma _x(\bm{k}) \cdot \Delta _{l\alpha}^{(m)}(\bm{k})=\Delta _{l\alpha}^{(m)}(-k_x,k_y,k_z)&= s_m \cdot \Delta _{l\alpha}^{(m)}(k_x,k_y,k_z),\\
\sigma _x(\bm{k})\cdot \sigma _x(\bm{k}\pri)\cdot V_{ll\pri}(\bm{k},\bm{k}\pri)=V_{ll\pri}&(-k_x,k_y,k_z,-k_x\pri ,k_y\pri ,k_z\pri) \notag\\
=V_{ll\pri}&(k_x,k_y,k_z,k_x\pri ,k_y\pri ,k_z\pri), 
\end{align}
where
$s_m$ is a phase factor of $\sigma _x$.
Now we restrict the region of integration to $k_x>0$, $k\pri_x>0$. Then, $I_{mn}$ is rewritten as 
\begin{align}
I_{mn}= 
\frac{1}{(2\pi)^6}&\sum _{ll\pri}\iint \frac{dS_{\bm{k} l}}{v_F(\bm{k},l)} \frac{dS_{\bm{k}\pri l\pri}}{v_F(\bm{k}\pri,l\pri)}
\notag \\ &\qquad \times \  \left( \Delta_{l\alpha}^{(m)}(\bm{k})\right)^*\tilde{V}_{ll\pri}(\bm{k},\bm{k}\pri)\Delta_{l\pri \alpha}^{(n)}(\bm{k}\pri), \\
\tilde{V}_{ll\pri}(\bm{k},\bm{k}\pri)&=\left[ (1+s_m s_n)+(s_m + s_n)\sigma _x(\bm{k}) \right]V_{ll\pri}(\bm{k},\bm{k}\pri).
\end{align}
When we put $m=xy$ and $n=yz$, for example, 
we obtain $s_{m}=-1,\ s_{n}=1$.
Then, $\tilde{V}_{ll\pri}(\bm{k},\bm{k}\pri)$ and hence $I_{mn}$ become zero. 
Also for other cases including $E_g$,  we generally find 
that $I_{mn}=0$ for $m\neq n$.
Therefore we need only a single basis function in 
to obtain the maximum eigenvalue $\lambda_{\alpha}$ for each $\alpha$.

\label{lastpage}
\clearpage

\end{document}